\documentclass[a4paper,fleqn,usenatbib]{mnras}

\usepackage[T1]{fontenc}
\usepackage{ae,aecompl}

\usepackage{graphicx}	% Including figure files
\usepackage{amsmath}	% Advanced maths commands
\usepackage{amssymb}	% Extra maths symbols

\def\msun{{\rm\,M_\odot}}

\def\gtsima{$\; \buildrel > \over \sim \;$}
\def\simgt{\lower.5ex\hbox{\gtsima}}

\def\msun{\hbox{M$_\odot$}}
\def\songmei2022{https://doi.org/10.48550/arXiv.2212.11034}

\title[The SMC metalllicity distribution]{An in-depth view of the metallicity distribution of the
Small Magellanic Cloud}

\author[Andr\'es E. Piatti]{
Andr\'es E. Piatti$^{1,2}$\thanks{E-mail: andres.piatti@fcen.uncu.edu.ar} \\
$^{1}$Instituto Interdisciplinario de Ciencias B\'asicas (ICB), CONICET-UNCUYO, 
Padre J. Contreras 1300, M5502JMA, Mendoza, Argentina\\
$^{2}$Consejo Nacional de Investigaciones Cient\'{\i}ficas y T\'ecnicas, Godoy Cruz 
2290, C1425FQB,  Buenos Aires, Argentina\\
}

\date{Accepted XXX. Received YYY; in original form ZZZ}

\pubyear{2022}

\begin{document}
\label{firstpage}
\pagerange{\pageref{firstpage}--\pageref{lastpage}}
\maketitle

\begin{abstract}
The spatial metallicity distribution of star clusters in the Small Magellanic Cloud (SMC) 
has recently been found to correlate as a V-shaped function with the semi-major axis of 
an elliptical framework proposed to assume a projected galaxy flattening. 
We report results on the impact that the use of such a framework 
can produce on our understanding of the SMC formation and its chemical enrichment. We show 
that clusters with similar semi-major axes are placed
at a very different distances from the SMC centre.
The recently claimed bimodal metallicity distribution of clusters projected on the
innermost SMC regions and the V-shaped metallicity gradient fade away when actual 
distances are used. Although a large dispersion prevails,
clusters older than $\sim$ 1 Gyr exhibit a shallow metallicity gradient, caused
by slightly different spatial distributions of clusters younger and 
older than $\sim$ 4 Gyr; the former being more centrally concentrated and having a
mean metallicity ([Fe/H]) $\sim$ 0.15 dex more metal-rich than that of older clusters. 
This metallicity gradient does not show any dependence with the position angle, except
for clusters placed beyond 11 kpc, which are located in the eastern side of the galaxy. 
\end{abstract} 


%%%%%%%%%%%%%%%%%%%%%%%%%%%%%%%%%%%%%%%%%%%

%%%%%%%%%%%%%%%%% BODY OF PAPER %%%%%%%%%%%%%%%%%

\section{Introduction}

The spatial metallicity distribution of the Small Magellanic Cloud (SMC)
has recently been reanalysed by \citet{debortolietal2022} from a 
compilation of metallicities of 57 star
clusters. They derived mean Ca\,II triplet metallicities for six SMC star
clusters, which were added to previous metallicity estimates derived using the 
same technique of other 51 star clusters, taken from their previous works or 
the literature. They showed that the metallicity estimates of these 57
star clusters are on the same scale. As far as we are aware, this is the largest
compilation of SMC star cluster metallicities.  From the analysis
of this spectroscopic dataset as a function of the projected distance, 
\citet{debortolietal2022} found that there would seem to be a 
bimodal metallicity distribution for star clusters located in the innermost
region of the SMC.  Apparently, both metal-rich and metal-poor star cluster 
groups do not exhibit 
any metallicity gradient, but considered together, they show a negative metallicity
gradient similar to that of field stars,  which show a unimodal distribution
throughout the SMC main body. For star clusters located beyond the
SMC innermost region, \citet{debortolietal2022} found a positive metallicity 
gradient.

The change in the slope of the metallicity gradient, from a negative to a positive value
while moving from the SMC centre outwards, was previously suggested by 
\citet{parisietal2009,parisietal2015,bicaetal2020,parisietal2022} and \citet{oliveiraetal2023},
among others, who successively enlarged the star cluster sample analysed. The recent work by
\citet{debortolietal2022} is based on these previous ones. Previously, \citet{p11b} analysed a 
statistically complete old SMC star cluster sample (age $\la$ 1 Gyr) and found no metallicity 
gradient, but a metallicity spread across the entire SMC body. \citet{williamsetal2022},
more recently, performed numerical simulations to describe the formation of the old SMC
star cluster population assuming a negligible metallicity gradient.

In this work we  in-depth revisit the analysis of the metallicity 
distribution for
star clusters located in the innermost SMC regions and of the V-shaped metallicity grandient. 
As can be seen, the actual metallicity distribution is a key piece of information to recover 
the SMC formation history and to properly trace its interaction with the Large Magellanic Cloud 
(LMC) and the Milky Way \citep{rubeleetal2018,massanaetal2022}. Since a V-shaped metallicity
gradient or a lack of any trend of the metallciity with the distance from the SMC centre
implies different channels for the SMC formation and evolution, we think that 
reconciling the
above discrepancies is of an important impact for our comprehensive knowledge in this 
field of research. In Section~2, we
dig up  the constrains of previous approaches that led them to conclude on the metallicity
bimodality and V-shaped metallicity gradient, while in Section~3 we describe actual facts
that will comprehensively help reconstructing the SMC formation and its interaction history.

\begin{figure}
\includegraphics[width=\columnwidth]{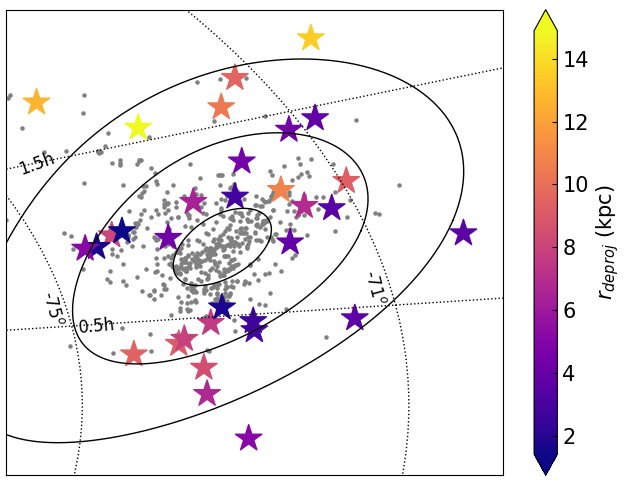}
\caption{Equal-area Hammer projection of the SMC in equatorial coordinates. Three ellipses
with semi-major axes of 1$\degr$, 3$\degr$ and 5$\degr$ are superimposed, respectively. 
Grey dots represent the star clusters catalogued in \citet{bicaetal2020}. Star symbols 
are coloured according to the actual star cluster distance to the SMC centre.}
\label{fig1}
\end{figure}

\section{Data analysis and discussion}

\citet{cetal01} used the angular distance to the SMC centre in right ascension and in 
declination and the line-of-sight depth of 12 old SMC star clusters to study their spatial 
distribution and the behaviour of their ages and metallicities with the position in the 
galaxy. They found that the SMC is a triaxial galaxy with the declination, the right ascension 
and the line-of-sight as the three axis with ratios of approximately 1:2:4. With the aim of 
mitigating the lack of accurate star cluster heliocentric distances in the literature, 
\citet{petal07b} introduced an elliptical framework that reflected more accurately the 
SMC flattening. The ellipse has a position angle of 54$\degr$ and a $b/a$ ratio of 1/2, with 
centre at RA = 00$^h$ 52$^m$ 45$^s$, Dec. = -72$\degr$ 49$\arcmin$ 43$\arcsec$ \citep{cetal01}.
They used the semi-major axis --parallel to the SMC main body-- as a meaningful indicator of the 
projected distance to the SMC centre. Thus, they assumed that cluster age and metallicity variations, 
if any, correlate much better with a pseudo-elliptical (projected) distance measured from the 
galaxy centre than with the radial distance, or distances defined along the right ascension or 
declination axes. For the sake of the reader, Fig.~\ref{fig1} shows as an example three
ellipses superimposed on the spatial distribution of the star clusters catalogued by 
\citet{bicaetal2020}. 

\begin{figure}
\includegraphics[width=\columnwidth]{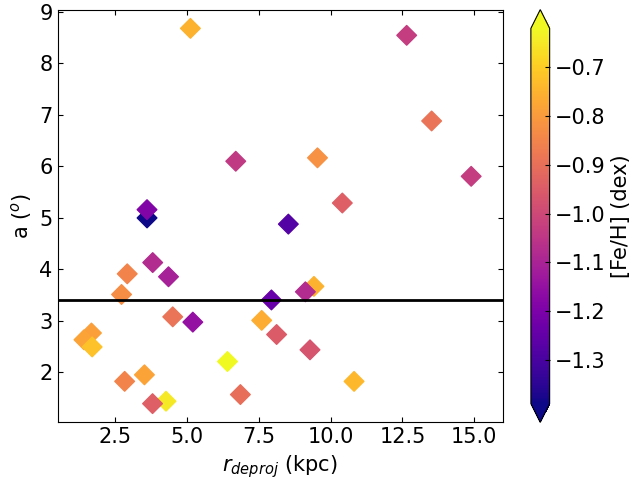}
\caption{Relation between the semi-major axis ($a$) and the computed deprojected
distances. Symbols refer to star clusters with different metallicity values. The black
horizontal line represent the boundary of the innermost SMC region ($a$ = 3.4$\degr$) 
adopted by \citet{debortolietal2022}. }
\label{fig2}
\end{figure}

Such an elliptical framework has been adopted by different authors when dealing with the
SMC metallicity gradient. For instance, \citet[][see also \citet{choudhuryetal2018}]{choudhuryetal2020} 
employed near-infrared photometry from the VISTA Survey of the Magellanic Clouds \citep[VMC,][]{cetal11} 
to map metallicity trends in the SMC using the elliptical framework of \citet{petal07b}. They found
that field stars show a unimodal metallicity distribution across the galaxy with a
shallow V-shaped metallicity gradient. As star clusters are considered, several works arrived to the
conclusion of a V-shaped metallicity gradient, in very good agreement with outcomes from the 
analysis of field stars \citep[e.g.,][and references therein]{bicaetal2020,parisietal2022,oliveiraetal2023}.
\citet{debortolietal2022} added to this picture of the SMC metallicty map the result of a bimodal
metallicity distribution for stars clusters located within the innermost regions of the galaxy
(semi-major axis $a$ $<$ 3.4$\degr$). 

The aforementioned framework does not consider the depth of the SMC, which is much more
extended than the size of the galaxy projected in the sky \citep{ripepietal2017,muravevaetal2018,graczyketal2020}.
In this sense, star clusters observed projected onto the innermost regions can be distributed
along the whole line-of-sight, so that their distances from the SMC centre can also be very 
different. This simple possibility has important consequences in our understanding of the SMC
formation process. Whether the SMC formed from a outside-in radial collapse \citep{pt1998}, 
from a major merger \citet{tb2009}, from a closed-box formation model \citep{dh98}, etc,
has its own implications in the conclusions that can be drawn from the observed
spatial metallicity distribution of star clusters.

\begin{figure}
\includegraphics[width=\columnwidth]{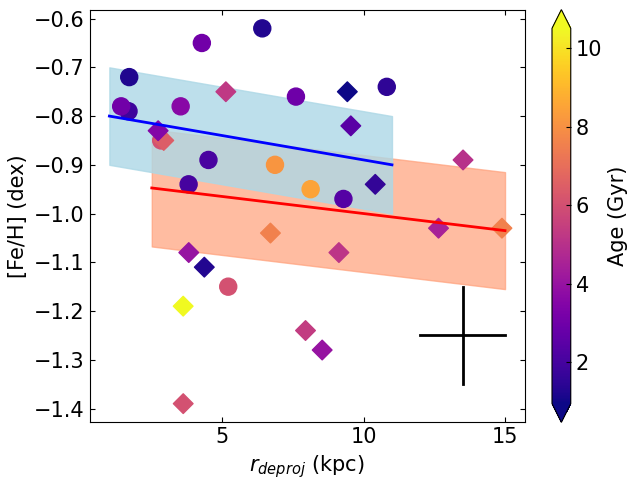}
\caption{Star cluster metallicity distribution as a function of their deprojected
distances. Coloured symbols refer to star cluster ages.  Circles and diamonds
represent star clusters with $a$ values smaller and larger than 3.4$\degr$ (see 
Fig.~\ref{fig2}), respectively. Blue (-0.010$\times$$r_{deproj}$ -0.790; rms=0.100) and 
red (-0.007$\times$$r_{deproj}$ -0.930; rms=0.120) lines with their shaded regions,
represent the resulting linear fits for each group, respectively. Typical error bars are
also indicated.}
\label{fig3}
\end{figure}

\begin{table}
\caption{SMC star cluster properties.}
\label{tab1}
\begin{tabular}{@{}lcccc}\hline
ID & $d$     & Ref. & [Fe/H] & Ref.  \\
     & (kpc) &      &  (dex) &       \\\hline

AM3         &  62.8$\pm$4.6 &  1 &  -0.75$\pm$0.11 & 4 \\
B99         &  ---          &    &  -0.84$\pm$0.04 & 5 \\ 
B168        &  61.9$\pm$2.1 &  2 &  -1.08$\pm$0.09 & 10 \\ 
BS121       &  ---          &    &  -0.66$\pm$0.07 & 6 \\
BS188       &  52.7$\pm$3.0 &  2 &  -0.94$\pm$0.13 & 10 \\ 
BS196       &  50.1$\pm$2.0 &  2 &  -0.89$\pm$0.08 &  10 \\  
H86-97      &  ---          &    &  -0.71$\pm$0.05 & 5 \\ 
HW31        &  ---          &    &  -1.12$\pm$0.37 & 8 \\ 
HW40        &  65.5$\pm$4.2 &  1 &  -0.78$\pm$0.05 & 5 \\  
HW41        &  ---          &    &  -0.96$\pm$0.36 & 8 \\ 
HW42        &  54.5$\pm$2.7 &  3 &  -0.95$\pm$0.42 & 8 \\ 
HW47        &  ---          &    &  -0.92$\pm$0.04 & 6 \\ 
HW56        &  53.5$\pm$1.2 &  2 &  -0.97$\pm$0.20 & 10 \\  
HW67        &  ---          &    &  -0.72$\pm$0.04 & 5 \\ 
HW84        &  ---          &    &  -0.91$\pm$0.05 & 6 \\ 
HW85        &  54.0$\pm$1.6 &  2 &  -0.82$\pm$0.14 &  10 \\ 
HW86        &  ---          &    &  -0.61$\pm$0.06 & 6 \\
IC1708      &  65.2$\pm$1.5 &  2 &  -1.11$\pm$0.17 & 10 \\  
K3          &  60.6$\pm$1.1 &  2 &  -0.85$\pm$0.03 & 5 \\  
K6          &  ---          &    &  -0.63$\pm$0.02 & 5 \\ 
K7          &  64.3$\pm$2.4 &  4 &  -0.83$\pm$0.06 & 7 \\ 
K8          &  69.8$\pm$2.3 &  2 &  -0.76$\pm$0.07 &  4 \\ 
K9          &  ---          &    &  -1.12$\pm$0.05 &  5 \\
K37         &  62.4$\pm$2.0 &  4 &  -0.79$\pm$0.11 &  5 \\  
K38         &  66.7$\pm$1.9 &  4 &  -0.65$\pm$0.18 & 8 \\  
K44         &  62.2$\pm$2.7 &  2 &  -0.78$\pm$0.03 & 7 \\ 
L1          &  56.9$\pm$1.0 &  2 &  -1.04$\pm$0.03 & 5 \\ 
L2          &  54.5$\pm$3.5 &  1 &  -1.28$\pm$0.09 & 4 \\ 
L3          &  53.5$\pm$3.4 &  1 &  -0.75$\pm$0.33 & 8 \\ 
L4          &  53.7$\pm$2.4 &  2 &  -1.08$\pm$0.04 &  6 \\ 
L5          &  ---          &    &  -1.25$\pm$0.05 &  6 \\ 
L6          &  54.9$\pm$2.3 &  2 &  -1.24$\pm$0.03 &  6 \\ 
L7          &  ---          &    &  -0.76$\pm$0.06 &  6 \\
L19         &  ---          &    &  -0.87$\pm$0.03 &  6 \\  
L27         &  ---          &    &  -1.14$\pm$0.06 &  6 \\ 
L32         &  ---          &    &  -0.96$\pm$0.04 &  7 \\  
L38         &  64.0$\pm$1.1 &  1 &  -1.39$\pm$0.03 &  7 \\  
L43         &  58.8$\pm$3.3 &  4 &  -0.94$\pm$0.03 &  7 \\  
L91         &  ---          &    &  -1.01$\pm$0.35 &  8 \\
L100        &  58.6$\pm$0.7 &  2 &  -0.89$\pm$0.14 & 10 \\
L106        &  ---          &    &  -0.88$\pm$0.06 &  6 \\ 
L108        &  ---          &    &  -1.05$\pm$0.05 &  6 \\   
L110        &  47.9$\pm$2.3 &  2 &  -1.03$\pm$0.05 &  6 \\  
L112        &  ---          &    &  -1.08$\pm$0.07 &  5 \\  
L113        &  50.5$\pm$1.7 &  2 &  -1.03$\pm$0.04 &  5 \\ 
L116        &  ---          &    &  -0.89$\pm$0.02 &  7 \\
NGC121      &  64.9$\pm$1.2 &  2 &  -1.19$\pm$0.12 &  9 \\
NGC152      &  61.9$\pm$5.4 &  1 &  -0.72$\pm$0.02 &  7 \\   
NGC339      &  57.6$\pm$4.1 &  2 &  -1.15$\pm$0.02 &  7 \\
NGC361      &  55.8$\pm$1.7 &  2 &  -0.90$\pm$0.03 &  7 \\    
NGC411      &  51.8$\pm$3.3 &  1 &  -0.74$\pm$0.04 &  7 \\  
NGC416      &  60.0$\pm$1.9 &  1 &  -0.85$\pm$0.04 &  7 \\  
NGC419      &  56.2$\pm$1.3 &  2 &  -0.62$\pm$0.02 &  7 \\  
NGC643      &  ---          &    &  -0.82$\pm$0.03 &  6 \\
OGLE-SMC113 &  ---          &    &  -0.80$\pm$0.07 &  5 \\\hline  

\end{tabular}

\noindent (1)\citet{dgb15}; (2)\citet{piatti2021f}; (3)\citet{piatti2022b}; 
(4)\citet{diasetal2022}; (5)\citet{parisietal2015}; (6)\citet{parisietal2009}; 
(7)\citet{parisietal2022}; (8)\citet{debortolietal2022}; (9)\citet{dh98}; 
(10)\citet{diasetal2021}.

\end{table}

With the aim of performing a more realistic analysis of the star cluster spatial metallicity 
distribution, we here introduce the actual distance of an star cluster to the SMC centre 
as an independent variable,  In order to compute star cluster distances to the SMC centre,
we made use of: i) the compilation of star clusters of \citet{debortolietal2022}; ii) the mean SMC 
heliocentric distance \citep[62.44$\pm$0.81 kpc;][]{graczyketal2020,dgb15};  and iii) an 
homogeneous compilation of accurate star cluster heliocentric distances. The 57 star clusters
in \citet{debortolietal2022} were used for comparison purposes of the resulting metallicity 
gradients by using their derived star cluster metallicities (see Table~\ref{tab1}). As for the 
star cluster distances, we 
based our analysis on the distance scale built from accurate heliocentric distance estimates
by \citet[][see Table~1]{piatti2021f}. To avoid repetition, we refer the reader to that work 
for details on the construction of that distance scale. Besides  the compilation of
distances in \citet{piatti2021f}, we included those in \citet{dgb15} 
(AM3, HW40, Lindsay~2, Lindsay~3, Lindsay~38, NGC~152, NGC~411, NGC~416), 
in \citet{diasetal2022} (Kron~7, Kron~37, Kron~38, Lindsay~43), and in \citet{piatti2022b} 
(HW~42), respectively.  All the added star cluster distances are put in the \citet{piatti2021f}'s
distance scale. We gathered  in total 32 star clusters with accurate distances and with
metallicity estimates in \citet{debortolietal2022}  (see Table~\ref{tab1}).

The mean SMC distance \citep[$R_{LMC}$=62.5 kpc][]{graczyketal2020}, the star cluster distances 
($d$, see Table~\ref{tab1}) and the angular distances of the
star clusters to the SMC centre ($a$) were used to compute deprojected distances ($r_{deproj}$)
as follows:\\

$r_{deproj} ^2= {R_{LMC}}^2 + d^2 - 2 \times R_{LMC} \times d \times cos(a)$\\

\noindent Fig.~\ref{fig1} shows the spatial distribution of the 32 star clusters painted with different 
colours according to their $r_{deproj}$ values. As can be seen, star clusters located close to an ellipse 
(e.g., $a$ $\approx$ 5$\degr$) have different deprojected distances ($r_{deproj}$ $\sim$ 3 - 
14 kpc), which confirms the suspicion that clusters at different distances to the SMC centre can 
have similar semi-major axis values. The uncovered behaviour is highlighted in Fig.~\ref{fig2}, 
which reveals the lack of linearity between the semi-major axis and the deprojected distance. 
Particularly, Fig.~\ref{fig2} shows that the innermost star clusters selected by \citet{debortolietal2022}
($a$ $<$ 3.4$\degr$) span deprojected distances from $\sim$ 1.4 up to 10.8 kpc. Therefore,
star clusters projected on to the innermost regions are not necessarily located close to the
SMC centre. This seems a
straightforward outcome, that is helpful to quantify in order to assess the level of
accuracy of the  interpretations of the spatial metallicity
distributions built using semi-major axes. The limited number of SMC star clusters with
accurate heliocentric distance estimates calls our attention of the need of  an effort to homogeneously 
determine distances for a large sample of clusters. Fortunately, there are ongoing observing campaigns
aimed at obtaining homogeneous data for the Magellanic Clouds \citep[e.g., see Table 1 of][]{maiaetal2019}.

From Fig.~\ref{fig2} we also found a unimodal metallicity distribution for star clusters located 
inside a volume of radius (deprojected distance) 3.5$\degr$ and estimated for them a 
mean metallicity of [Fe/H] = -0.80$\pm$0.05 dex. In consequence, the metallicity bimodality 
found by \citet{debortolietal2022} rather seems to reflect the superposition of star clusters 
with different metallicity values placed across the entire SMC extent along the innermost 
line-of-sight. On the other hand, a bimodal metallicity distribution in the innermost
region of the SMC, with a metal-poor ([Fe/H]$\sim$-1.15 dex) and a metal-rich 
([Fe/H]$\sim$-0.80 dex) peak, respectively, would imply two different star cluster formation 
epochs that took place only in the inner SMC body, while the outer SMC body kept without
noticing them. The SMC is a relative low-mass galaxy 
\citep[total mass $\sim$ 2$\times$10$^9$$\msun$,][]{stanimirovicetal2004}, in which
star clusters and field stars have evolved synchronically over time \citep{hz04,pg13,p15}, 
and where interactions with the LMC has triggered stellar formation throughout the entire
galaxy body \citep{p11b,p12a,rubeleetal2018}. In addition, unimodal spatial metallicity
distributions have been observed for field stars \citep{choudhuryetal2018,choudhuryetal2020}.

Fig.~\ref{fig3} shows the resulting metallicity gradient. Points are coloured
according to the  star cluster ages, which were taken from \citet{bicaetal2020} for
uniformity purposes. Fig.~\ref{fig3} suggests 
that an overall metallicity dispersion prevails
for star clusters older than $\sim$ 1 Gyr. Nevertheless, Fig.~\ref{fig3} also hints at star 
clusters younger than $\sim$ 4 Gyr are found distributed from the SMC centre out to $\sim$ 11 
kpc and with a mean [Fe/H] $\sim$ -0.85 dex; while older star clusters are distributed from 
$\sim$ 3 kpc out to outermost SMC regions and with a mean [Fe/H] $\sim$ -1.0 dex. This slight 
mean metallicty offset ($\Delta$[Fe/H]=0.15 dex) between younger and older star clusters, that 
in turn have slight different spatial distributions, tell us about a subtle metallicity gradient 
of $\sim$ -0.010$\pm$0.015 dex/kpc. Such a small metallicity gradient arises because the younger 
the star clusters (the slightly more centrally concentrated), the more metal-rich they are, within 
the overall metallicity dispersion.  For completeness purposes, we split the star cluster sample
in two groups with $a$ values smaller and larger than 3.4$\degr$ (circle and diamond in 
Fig.~\ref{fig3}), respectively. and performed linear regressions between their metallicities and
deprojected distances. Fig.~\ref{fig3} shows the results of these fits.
As can be seen, we did not detect any
V-shaped trend of the metallicity with the distance from the SMC centre, as shown by 
\citet[][and references therein]{debortolietal2022}. However, Fig.~\ref{fig2}
shows that star clusters with $a$ values between $\sim$ 3.0$\degr$ and 5.5$\degr$ are more metal-poor
than those outside that $a$ range. Therefore, we conclude that the V-shaped gradient is not
real, but a trend that appears when using  projected distances as an independent variable.
Likewise, field stars seem to share a shallow metallicity
gradient with a large dispersion \citep[see, e.g.,][and references therein]{mucciarellietal2023}. 

\begin{figure}
\includegraphics[width=\columnwidth]{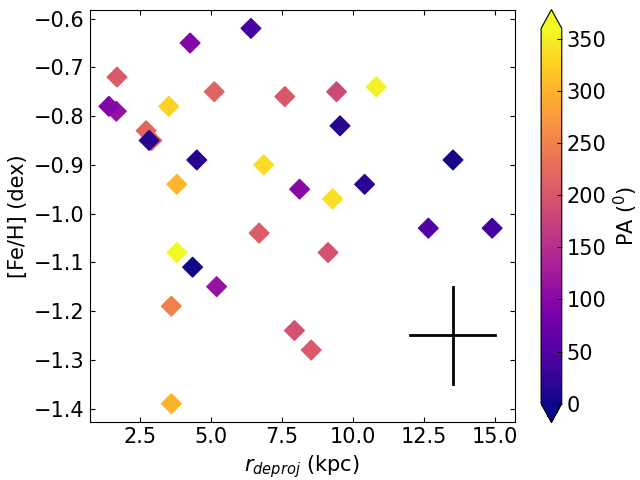}
\caption{Same as Fig.~\ref{fig3}, with coloured symbols
referring to Position Angle (see text for details).}
\label{fig4}
\end{figure}

We additionally examined whether the described metallicity gradient has any dependence
with the position angle (PA) measured eastward from the SMC major axis. We assumed that
the SMC major axis is rotated 54$\degr$ anticlockwise from the North. The PAs in this
rotated system were computed using the \texttt{positionAngle} routine from 
\texttt{PyAstronomy}\footnote{https://github.com/sczesla/PyAstronomy} \citep[PyA,][]{pya} 
and the observed distances in the sky to the SMC centre in RA ($x_0$) and
Dec ($y_0$), respectively, as follows:\\

$x_0$ = -(RA - RA$_{SMC}$)  cos(Dec) cos(PA$_{SMC}$) + (Dec - Dec$_{SMC}$) sin(PA$_{SMC}$),\\

$y_0$ =  (RA - RA$_{SMC}$)  cos(Dec) sin(PA$_{SMC}$) + (Dec - Dec$_{SMC}$) cos(PA$_{SMC}$).\\

\noindent Fig.~\ref{fig4} shows that there is not any dependency with the PA, with the sole
exception of the three farthest star clusters (BS196, Lindsay~110, Lindsay~113; $r_{deproj}$ $>$
11 kpc) located in the SMC eastern side. Their heliocentric distances ($\sim$ 47-50 kpc) are akin to 
that of the Magellanic Bridge, so that their positions reflect their motions
towards de LMC. Indeed, \citet{piatti2021b} fitted the rotation disc that best resembles the observed
motions of SMC star clusters and found that it is kinematically synchronised with that of
field red giants, showing some velocity stretching towards the LMC. For the sake of the
reader, if we considered the
different SMC sectors, namely: Northern bridge (-15 $\la$ PA ($\degr$) $\la$ 40); Bridge 
(40 $\la$ PA ($\degr$) $\la$ 90); West-halo (160 $\la$ PA ($\degr$) $\la$ 270); and 
Counter-bridge (300 $\la$ PA ($\degr$) $\la$ 345) \citep{diasetal2022}, we would find from Fig.~\ref{fig4} no different gradients.

\section{conclusions}

Since recent years, the spatial distribution of the SMC has been described as a V-shaped
function from its centre outwards, with a negative slope for the innermost region of the galaxy, 
and a positive one for the outer galaxy body. As far as we are aware, this kind of correlation of 
the metallicity with the position in a galaxy has only been observed in the SMC. Because of the
important implications of the knowledge of the observed spatial metallicity distribution in our 
understanding of the galaxy formation process, its chemical enrichment and the interaction with 
the LMC and the Milky Way, we revisited the construction of the SMC star cluster spatial metallicity
distribution. We found from a limited sample of star clusters that needs to be enlarged
for a larger coverage in deprojected distances that:\\

$\bullet$ The elliptical framework adopted to trace the spatial metallicity distribution 
\citep{petal07b}, by using the so-called semi-major axis as an independent variable, and from which
the V-shaped gradient arises, misleads
the star cluster positions. Star clusters with a similar semi-major axis are found to be located 
at very different distances from the SMC centre, in some cases by more than 10 kpc apart.\\

$\bullet$ The bimodal metallicty distribution for the innermost SMC region 
would seem to be caused by the consideration of projected distances instead
of the actual ones. Outer and inner star clusters that, according to their 3D positions have
different metallicities, are seen along the same line-of-sight.\\

$\bullet$ The star clusters (ages $>$  1 Gyr) show a shallow metallicity gradient (-0.01 dex/kpc) 
as a function of their deprojected distances to the SMC centre.  This gradient is caused by 
slightly different spatial distributions of star clusters younger and older than $\sim$ 4 Gyr. 
The younger ones are more centrally concentrated and have a mean metallicity ([Fe/H]) 
$\sim$ 0.15 dex more metal-rich than those older. Nevertheless, a overall metallicity dispersion
prevails.\\

$\bullet$ The spatial metallicity distribution of star clusters analysed in this work would 
not seem to  show any dependence
with the position angle, although there is some hint of star clusters placed beyond 11 kpc 
from the SMC centre to be located in the eastern side of the galaxy. Such a somehow spatial
asymmetry witnesses the interaction between both Magellanic Clouds, with SMC star clusters
moving towards the LMC.

\section*{Acknowledgements}
We thank the referee for the thorough reading of the manuscript and timely suggestions to improve it.
%This work has made use of data from the European Space Agency (ESA) mission
%{\it Gaia} (\url{https://www.cosmos.esa.int/gaia}), processed by the {\it Gaia}
%Data Processing and Analysis Consortium (DPAC,
%\url{https://www.cosmos.esa.int/web/gaia/dpac/consortium}). Funding for the DPAC
%has been provided by national institutions, in particular the institutions
%participating in the {\it Gaia} Multilateral Agreement.

%This research made use of Astropy, a community-developed
%core Python package for Astronomy.

\section{Data availability}

Data used in this work are available upon request to the author.

%\newpage


%\bibliographystyle{mnras}
%\bibliography{paper} % if your bibtex file is called paper.bib

\begin{thebibliography}{}
\makeatletter
\relax
\def\mn@urlcharsother{\let\do\@makeother \do\$\do\&\do\#\do\^\do\_\do\%\do\~}
\def\mn@doi{\begingroup\mn@urlcharsother \@ifnextchar [ {\mn@doi@}
  {\mn@doi@[]}}
\def\mn@doi@[#1]#2{\def\@tempa{#1}\ifx\@tempa\@empty \href
  {http://dx.doi.org/#2} {doi:#2}\else \href {http://dx.doi.org/#2} {#1}\fi
  \endgroup}
\def\mn@eprint#1#2{\mn@eprint@#1:#2::\@nil}
\def\mn@eprint@arXiv#1{\href {http://arxiv.org/abs/#1} {{\tt arXiv:#1}}}
\def\mn@eprint@dblp#1{\href {http://dblp.uni-trier.de/rec/bibtex/#1.xml}
  {dblp:#1}}
\def\mn@eprint@#1:#2:#3:#4\@nil{\def\@tempa {#1}\def\@tempb {#2}\def\@tempc
  {#3}\ifx \@tempc \@empty \let \@tempc \@tempb \let \@tempb \@tempa \fi \ifx
  \@tempb \@empty \def\@tempb {arXiv}\fi \@ifundefined
  {mn@eprint@\@tempb}{\@tempb:\@tempc}{\expandafter \expandafter \csname
  mn@eprint@\@tempb\endcsname \expandafter{\@tempc}}}

\bibitem[\protect\citeauthoryear{{Bica}, {Westera}, {Kerber}, {Dias}, {Maia},
  {Santos}, {Barbuy}  \& {Oliveira}}{{Bica} et~al.}{2020}]{bicaetal2020}
{Bica} E.,  {Westera} P.,  {Kerber} L. d.~O.,  {Dias} B.,  {Maia} F.,  {Santos}
  Jo{\~a}o F.~C. J.,  {Barbuy} B.,   {Oliveira} R. A.~P.,  2020, \mn@doi [\aj]
  {10.3847/1538-3881/ab6595}, \href
  {https://ui.adsabs.harvard.edu/abs/2020AJ....159...82B} {159, 82}

\bibitem[\protect\citeauthoryear{{Choudhury}, {Subramaniam}, {Cole}  \&
  {Sohn}}{{Choudhury} et~al.}{2018}]{choudhuryetal2018}
{Choudhury} S.,  {Subramaniam} A.,  {Cole} A.~A.,   {Sohn} Y.-J.,  2018,
  \mn@doi [\mnras] {10.1093/mnras/sty087}, \href
  {http://adsabs.harvard.edu/abs/2018MNRAS.475.4279C} {475, 4279}

\bibitem[\protect\citeauthoryear{{Choudhury} et~al.,}{{Choudhury}
  et~al.}{2020}]{choudhuryetal2020}
{Choudhury} S.,  et~al., 2020, \mn@doi [\mnras] {10.1093/mnras/staa2140}, \href
  {https://ui.adsabs.harvard.edu/abs/2020MNRAS.497.3746C} {497, 3746}

\bibitem[\protect\citeauthoryear{{Cioni} et~al.,}{{Cioni}
  et~al.}{2011}]{cetal11}
{Cioni} M.-R.~L.,  et~al., 2011, \mn@doi [\aap] {10.1051/0004-6361/201016137},
  527, A116

\bibitem[\protect\citeauthoryear{{Crowl}, {Sarajedini}, {Piatti}, {Geisler},
  {Bica}, {Clari{\'a}}  \& {Santos}}{{Crowl} et~al.}{2001}]{cetal01}
{Crowl} H.~H.,  {Sarajedini} A.,  {Piatti} A.~E.,  {Geisler} D.,  {Bica} E.,
  {Clari{\'a}} J.~J.,   {Santos} Jr. J.~F.~C.,  2001, \mn@doi [\aj]
  {10.1086/321128}, 122, 220

\bibitem[\protect\citeauthoryear{{Czesla}, {Schr{\"o}ter}, {Schneider},
  {Huber}, {Pfeifer}, {Andreasen}  \& {Zechmeister}}{{Czesla}
  et~al.}{2019}]{pya}
{Czesla} S.,  {Schr{\"o}ter} S.,  {Schneider} C.~P.,  {Huber} K.~F.,  {Pfeifer}
  F.,  {Andreasen} D.~T.,   {Zechmeister} M.,  2019, {PyA: Python
  astronomy-related packages} (\mn@eprint {ascl} {1906.010})

\bibitem[\protect\citeauthoryear{{Da Costa} \& {Hatzidimitriou}}{{Da Costa} \&
  {Hatzidimitriou}}{1998}]{dh98}
{Da Costa} G.~S.,  {Hatzidimitriou} D.,  1998, \mn@doi [\aj] {10.1086/300340},
  115, 1934

\bibitem[\protect\citeauthoryear{{De Bortoli}, {Parisi}, {Bassino}, {Geisler},
  {Dias}, {Gimeno}, {Angelo}  \& {Mauro}}{{De Bortoli}
  et~al.}{2022}]{debortolietal2022}
{De Bortoli} B.~J.,  {Parisi} M.~C.,  {Bassino} L.~P.,  {Geisler} D.,  {Dias}
  B.,  {Gimeno} G.,  {Angelo} M.~S.,   {Mauro} F.,  2022, \mn@doi [\aap]
  {10.1051/0004-6361/202243762}, \href
  {https://ui.adsabs.harvard.edu/abs/2022A&A...664A.168D} {664, A168}

\bibitem[\protect\citeauthoryear{{Dias} et~al.,}{{Dias}
  et~al.}{2021}]{diasetal2021}
{Dias} B.,  et~al., 2021, \mn@doi [\aap] {10.1051/0004-6361/202040015}, \href
  {https://ui.adsabs.harvard.edu/abs/2021A&A...647L...9D} {647, L9}

\bibitem[\protect\citeauthoryear{{Dias} et~al.,}{{Dias}
  et~al.}{2022}]{diasetal2022}
{Dias} B.,  et~al., 2022, \mn@doi [\mnras] {10.1093/mnras/stac259}, \href
  {https://ui.adsabs.harvard.edu/abs/2022MNRAS.512.4334D} {512, 4334}

\bibitem[\protect\citeauthoryear{{Graczyk} et~al.,}{{Graczyk}
  et~al.}{2020}]{graczyketal2020}
{Graczyk} D.,  et~al., 2020, \mn@doi [\apj] {10.3847/1538-4357/abbb2b}, \href
  {https://ui.adsabs.harvard.edu/abs/2020ApJ...904...13G} {904, 13}

\bibitem[\protect\citeauthoryear{{Harris} \& {Zaritsky}}{{Harris} \&
  {Zaritsky}}{2004}]{hz04}
{Harris} J.,  {Zaritsky} D.,  2004, \mn@doi [\aj] {10.1086/381953}, 127, 1531

\bibitem[\protect\citeauthoryear{{Maia} et~al.,}{{Maia}
  et~al.}{2019}]{maiaetal2019}
{Maia} F.~F.~S.,  et~al., 2019, \mn@doi [\mnras] {10.1093/mnras/stz369}, \href
  {http://adsabs.harvard.edu/abs/2019MNRAS.484.5702M} {484, 5702}

\bibitem[\protect\citeauthoryear{{Massana} et~al.,}{{Massana}
  et~al.}{2022}]{massanaetal2022}
{Massana} P.,  et~al., 2022, \mn@doi [\mnras] {10.1093/mnrasl/slac030}, \href
  {https://ui.adsabs.harvard.edu/abs/2022MNRAS.513L..40M} {513, L40}

\bibitem[\protect\citeauthoryear{{Mucciarelli}, {Minelli}, {Bellazzini},
  {Lardo}, {Romano}, {Origlia}  \& {Ferraro}}{{Mucciarelli}
  et~al.}{2023}]{mucciarellietal2023}
{Mucciarelli} A.,  {Minelli} A.,  {Bellazzini} M.,  {Lardo} C.,  {Romano} D.,
  {Origlia} L.,   {Ferraro} F.~R.,  2023, \mn@doi [\aap]
  {10.1051/0004-6361/202245133}, \href
  {https://ui.adsabs.harvard.edu/abs/2023A&A...671A.124M} {671, A124}

\bibitem[\protect\citeauthoryear{{Muraveva} et~al.,}{{Muraveva}
  et~al.}{2018}]{muravevaetal2018}
{Muraveva} T.,  et~al., 2018, \mn@doi [\mnras] {10.1093/mnras/stx2514}, \href
  {https://ui.adsabs.harvard.edu/abs/2018MNRAS.473.3131M} {473, 3131}

\bibitem[\protect\citeauthoryear{{Oliveira} et~al.,}{{Oliveira}
  et~al.}{2023}]{oliveiraetal2023}
{Oliveira} R.~A.~P.,  et~al., 2023, \mn@doi [arXiv e-prints]
  {10.48550/arXiv.2306.05503}, \href
  {https://ui.adsabs.harvard.edu/abs/2023arXiv230605503O} {p. arXiv:2306.05503}

\bibitem[\protect\citeauthoryear{{Pagel} \& {Tautvaisiene}}{{Pagel} \&
  {Tautvaisiene}}{1998}]{pt1998}
{Pagel} B.~E.~J.,  {Tautvaisiene} G.,  1998, \mn@doi [\mnras]
  {10.1046/j.1365-8711.1998.01792.x}, \href
  {https://ui.adsabs.harvard.edu/abs/1998MNRAS.299..535P} {299, 535}

\bibitem[\protect\citeauthoryear{{Parisi}, {Grocholski}, {Geisler},
  {Sarajedini}  \& {Clari{\'a}}}{{Parisi} et~al.}{2009}]{parisietal2009}
{Parisi} M.~C.,  {Grocholski} A.~J.,  {Geisler} D.,  {Sarajedini} A.,
  {Clari{\'a}} J.~J.,  2009, \mn@doi [\aj] {10.1088/0004-6256/138/2/517}, \href
  {http://adsabs.harvard.edu/abs/2009AJ....138..517P} {138, 517}

\bibitem[\protect\citeauthoryear{{Parisi}, {Geisler}, {Clari{\'a}},
  {Villanova}, {Marcionni}, {Sarajedini}  \& {Grocholski}}{{Parisi}
  et~al.}{2015}]{parisietal2015}
{Parisi} M.~C.,  {Geisler} D.,  {Clari{\'a}} J.~J.,  {Villanova} S.,
  {Marcionni} N.,  {Sarajedini} A.,   {Grocholski} A.~J.,  2015, \mn@doi [\aj]
  {10.1088/0004-6256/149/5/154}, \href
  {https://ui.adsabs.harvard.edu/abs/2015AJ....149..154P} {149, 154}

\bibitem[\protect\citeauthoryear{{Parisi}, {Gramajo}, {Geisler}, {Dias},
  {Clari{\'a}}, {Da Costa}  \& {Grebel}}{{Parisi}
  et~al.}{2022}]{parisietal2022}
{Parisi} M.~C.,  {Gramajo} L.~V.,  {Geisler} D.,  {Dias} B.,  {Clari{\'a}}
  J.~J.,  {Da Costa} G.,   {Grebel} E.~K.,  2022, \mn@doi [\aap]
  {10.1051/0004-6361/202142597}, \href
  {https://ui.adsabs.harvard.edu/abs/2022A&A...662A..75P} {662, A75}

\bibitem[\protect\citeauthoryear{{Piatti}}{{Piatti}}{2011}]{p11b}
{Piatti} A.~E.,  2011, \mn@doi [\mnras] {10.1111/j.1745-3933.2011.01145.x},
  418, L69

\bibitem[\protect\citeauthoryear{{Piatti}}{{Piatti}}{2012}]{p12a}
{Piatti} A.~E.,  2012, \mn@doi [\mnras] {10.1111/j.1365-2966.2012.20684.x},
  422, 1109

\bibitem[\protect\citeauthoryear{{Piatti}}{{Piatti}}{2015}]{p15}
{Piatti} A.~E.,  2015, \mn@doi [\mnras] {10.1093/mnras/stv1179}, \href
  {http://adsabs.harvard.edu/abs/2015MNRAS.451.3219P} {451, 3219}

\bibitem[\protect\citeauthoryear{{Piatti}}{{Piatti}}{2021a}]{piatti2021f}
{Piatti} A.~E.,  2021a, \mn@doi [\mnras] {10.1093/mnras/stab2796}, \href
  {https://ui.adsabs.harvard.edu/abs/2021MNRAS.508.3748P} {508, 3748}

\bibitem[\protect\citeauthoryear{{Piatti}}{{Piatti}}{2021b}]{piatti2021b}
{Piatti} A.~E.,  2021b, \mn@doi [\aap] {10.1051/0004-6361/202140643}, \href
  {https://ui.adsabs.harvard.edu/abs/2021A&A...650A..52P} {650, A52}

\bibitem[\protect\citeauthoryear{{Piatti}}{{Piatti}}{2022}]{piatti2022b}
{Piatti} A.~E.,  2022, \mn@doi [Research Notes of the American Astronomical
  Society] {10.3847/2515-5172/acabc6}, \href
  {https://ui.adsabs.harvard.edu/abs/2022RNAAS...6..271P} {6, 271}

\bibitem[\protect\citeauthoryear{{Piatti} \& {Geisler}}{{Piatti} \&
  {Geisler}}{2013}]{pg13}
{Piatti} A.~E.,  {Geisler} D.,  2013, \mn@doi [\aj]
  {10.1088/0004-6256/145/1/17}, 145, 17

\bibitem[\protect\citeauthoryear{{Piatti}, {Sarajedini}, {Geisler}, {Gallart}
  \& {Wischnjewsky}}{{Piatti} et~al.}{2007}]{petal07b}
{Piatti} A.~E.,  {Sarajedini} A.,  {Geisler} D.,  {Gallart} C.,
  {Wischnjewsky} M.,  2007, \mn@doi [\mnras]
  {10.1111/j.1365-2966.2007.12439.x}, 382, 1203

\bibitem[\protect\citeauthoryear{{Ripepi} et~al.,}{{Ripepi}
  et~al.}{2017}]{ripepietal2017}
{Ripepi} V.,  et~al., 2017, \mn@doi [\mnras] {10.1093/mnras/stx2096}, \href
  {https://ui.adsabs.harvard.edu/abs/2017MNRAS.472..808R} {472, 808}

\bibitem[\protect\citeauthoryear{{Rubele} et~al.,}{{Rubele}
  et~al.}{2018}]{rubeleetal2018}
{Rubele} S.,  et~al., 2018, \mn@doi [\mnras] {10.1093/mnras/sty1279}, \href
  {http://adsabs.harvard.edu/abs/2018MNRAS.478.5017R} {478, 5017}

\bibitem[\protect\citeauthoryear{{Stanimirovi{\'c}}, {Staveley-Smith}  \&
  {Jones}}{{Stanimirovi{\'c}} et~al.}{2004}]{stanimirovicetal2004}
{Stanimirovi{\'c}} S.,  {Staveley-Smith} L.,   {Jones} P.~A.,  2004, \mn@doi
  [\apj] {10.1086/381869}, \href
  {https://ui.adsabs.harvard.edu/abs/2004ApJ...604..176S} {604, 176}

\bibitem[\protect\citeauthoryear{{Tsujimoto} \& {Bekki}}{{Tsujimoto} \&
  {Bekki}}{2009}]{tb2009}
{Tsujimoto} T.,  {Bekki} K.,  2009, \mn@doi [\apjl]
  {10.1088/0004-637X/700/2/L69}, \href
  {https://ui.adsabs.harvard.edu/abs/2009ApJ...700L..69T} {700, L69}

\bibitem[\protect\citeauthoryear{{Williams}, {Bekki}  \& {McKenzie}}{{Williams}
  et~al.}{2022}]{williamsetal2022}
{Williams} M.~L.,  {Bekki} K.,   {McKenzie} M.,  2022, \mn@doi [\mnras]
  {10.1093/mnras/stab3638}, \href
  {https://ui.adsabs.harvard.edu/abs/2022MNRAS.512.4086W} {512, 4086}

\bibitem[\protect\citeauthoryear{{de Grijs} \& {Bono}}{{de Grijs} \&
  {Bono}}{2015}]{dgb15}
{de Grijs} R.,  {Bono} G.,  2015, \mn@doi [\aj] {10.1088/0004-6256/149/6/179},
  \href {https://ui.adsabs.harvard.edu/abs/2015AJ....149..179D} {149, 179}

\makeatother
\end{thebibliography}

%to be uncommented before sending to editor
%\input{paper.bbl}

%\begin{thebibliography}{99}
%\end{thebibliography}

%%%%%%%%%%%%%%%%%%%%%%%%%%%%%%%%%%%%%%%%%%%%%%%%%%
%%%%%%%%%%%%%%%% APPENDICES %%%%%%%%%%%%%%%%%%%%%

% Don't change these lines
\bsp	% typesetting comment
\label{lastpage}
\end{document}